# Call Admission Control and Traffic Modeling for Integrated Macrocell/Femtocell Networks

Mostafa Zaman Chowdhury and Yeong Min Jang
Department of Electronics Engineering, Kookmin University, Seoul 136-702, Korea.
E-mail: mzceee@yahoo.com, yjang@kookmin.ac.kr

**Abstract:** Dense femtocells and the integration of these femtocells with the macrocell are the ultimate goal of the femtocellular network deployment. Integrated macrocell/femtocell networks surely able to provide high data rate for the indoor users as well as able to offload huge traffic from the macrocellular networks to femtocellular networks. Efficient handling of handover calls is the key for the successful macrocell/femtocell integration. An appropriate traffic model for the integrated macrocell/femtocell networks is also needed for the performance analysis measurement. In this paper we presented a call admission control process and a traffic model for the integrated macrocell/femtocell networks. The numerical and simulation results show the important of the integrated macrocell/femtocell network and the performance improvement of the proposed schemes.

## I. Introduction

The data rate, quality of service (QoS), and cost are the most important parameters for the current and future wireless communications. The femtocellular network [1]-[4], one of the most promising technologies to meet the demand of the tremendous increasing wireless capacity by various wireless applications for the future wireless communications. Femtocellular network has several advantages compared to the other indoor wireless networks. Femtocells operate in the spectrum licensed for cellular service providers. The key feature of the femtocell technology is that users require no new equipment (UE). Deployment cost of the femtocell is very small and it provides high data rate. Therefore, the deployment of femtocells in the large scale and the integration of these femtocells with the overlaid macrocell are the ultimate goal for this technology. A well-designed macrocell/femtocell integrated network can divert huge amount of traffic from the congested and expensive macrocellular networks to femtocellular networks.

Fig. 1 shows an example of co-existing femtocellular and macrocellular networks. Thousands of femto access points (FAPs) are deployed under the macrocellular network coverage. In the overlaid macrocell coverage area, femtocell-to-femtocell, femtocell-to-macrocell, and macrocell-to-femtocell handovers are occurred due to the deployment of femtocells. The frequency of these handovers is increased as the density of femtocells is increased. The efficient femtocell-to-femtocell and femtocell-to-macrocell handovers result in seamlessly movement of the users. Even though the macrocell-to-femtocell handover is not essential for the seamlessly movement, but efficient handling of this handover can reduce huge traffic load of macrocellular networks by transferring the calls to femtocells.

The integration of the dense scale femtocells with the macrocellular network suffers from several challenges [1]-[3], [5]. The handover management is one challenging issue among several challenges. An effective call admission control (CAC) policy can manage the handover calls efficiently. Therefore, we suggest a CAC policy to handle various types of calls. The proposed CAC does not differentiate between the new originating calls and handover calls for the femtocellular networks due to available resources in femtocellular networks. The CAC provides higher priority for the handover calls in the overlaid macrocellular network by offering QoS adaptation provision [6], [7]. The QoS adaptation provision is only available to accept the handover calls in macrocellular network. Thus, the macrocellular network can accept huge number of handover calls that are generated due to the femtocells and the neighbor macrocells.

An appropriate traffic model is required to measure the performances. Therefore, we also propose a traffic model for the integrated macrocell/femtocell scenario. The existing traffic model should be modified to apply for the macrocell/femtocell integrated networks. The proposed traffic model for the macrocell/femtocell integrated networks is useful to analysis the performance of the macrocell/femtocell integrated networks.

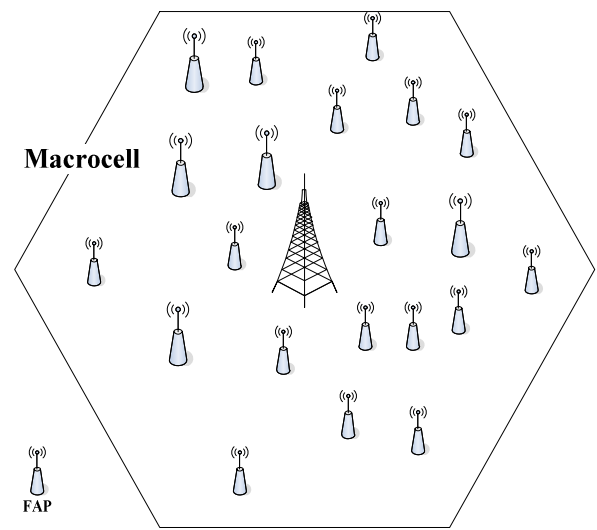

**Fig. 1.** An example of dense femtocells co-existing with macrocell.

The rest of this paper is organized as follows. CAC policies are provided in Section II. In Section III, we derive the detail traffic model and queuing analysis for the macrocell/femtocell integrated networks. Performance evaluation results of the proposed schemes are presented and compared in Section IV. Finally, Section V concludes our work.

## II. Proposed CAC Policy

An effective CAC can play a vital role to maximize the resource utilization for any kind of wireless networks. The macrocell resources are very valuable and limited. Thus for the macrocell/femtocell integrated networks, special attention should be taken to efficiently control the admission of various traffic calls inside the macrocell coverage area. The main goal of our proposed scheme is to transfer macrocell calls to femtocellular networks as many as possible. We divide the proposed CAC into three parts. The first one for the new originating calls, the second one for the calls that are originally connected with the macrocellular BS, and the last one for the calls that are originally connected with the FAPs. We offer the bandwidth degradation [6], [7] policy of the QoS adaptive multimedia traffic to accommodate more number of femtocell-to-macrocell and macrocell-to-macrocell handover calls. The existing QoS adaptive multimedia traffic in overlaid macrocellular network releases $C_{release,m}$ amount of bandwidth to accept the handover calls in the macrocellular network. This releasable amount depends on the number of running QoS adaptive multimedia calls and their maximum level of allowable QoS degradation and total number of existing calls in the macrocellular network. Suppose, $\beta_{r,m}$ and $\beta_{min,m}$, respectively, the requested and the minimum required bandwidth for each of the requesting calls in macrocell. Hence, each of the QoS adaptive calls can release maximum $\beta_{r,m} - \beta_{min,m}$ amount of bandwidth to accept a handover call in the macrocell system. If $C$ and $C_{occupied,m}$ are, respectively, the macrocell system bandwidth capacity and the occupied bandwidth by the existing macrocell calls, then the available empty bandwidth, $C_{availale,m}$ in the macrocellular network is $C - C_{occupied,m}$.

Fig. 2 shows the proposed CAC policy. Whenever a call is arrived, the CAC checks the call type. The priority for the new call is lower than that of a handover call for the resources of macrocellular networks. Whenever a new call is arrived, the CAC initially checks whether the femtocell coverage is available or not. If the femtocell coverage is available then FAP is the first choice to connect a call. A FAP accepts a new originating call if resource in the FAP is available. If the above condition is not satisfied, then the call tries to connect with the overlaid macrocellular network. The macrocell system does not allow QoS degradation policy to accept any new originating call. If a handover call is arrived, then the CAC checks whether it is macrocell-to-femtocell handover or femtocell-to-femtocell or femtocell-to-macrocell handover call. For a macrocell-to-femtocell call, whenever the moving MS detects signal from a FAP, the CAC policy checks the resource availability in the target FAP (T-FAP).

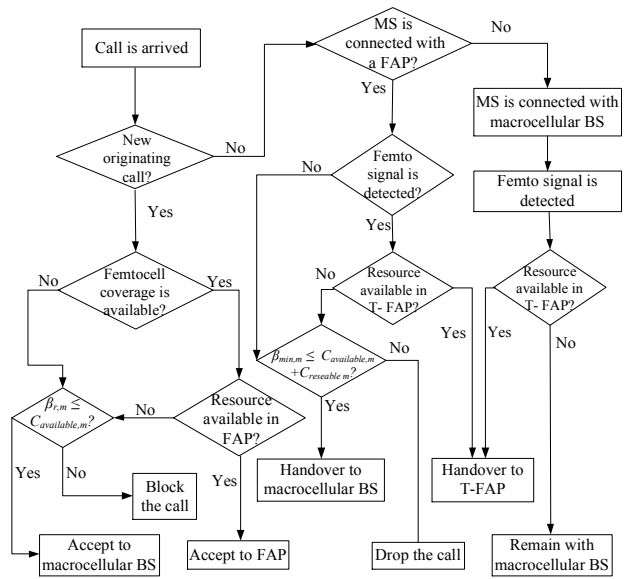

**Fig. 2.** Proposed CAC policy.

For a call of type femtocell-to-femtocell or femtocell-to-macrocell handover, whenever the signal level from the serving FAP (S-FAP) is going down, the MS initiates handover to other femtocell or overlaid macrocell. If another FAP is not available for handover or resource is not available in the T-FAP, the call tries to connect with the macrocellular network. If the empty resource in the macrocell system is not enough to accept the call, the CAC policy allows releasing of some bandwidth from the existing calls by degrading the QoS level of them. The CAC policy also permits the reduction of required bandwidth from $\beta_{r,m}$ to $\beta_{min,m}$ for a handover call request. If the minimum required bandwidth $\beta_{min,m}$ is not available in the macrocell system after releasing of some bandwidth from the existing calls, then the call is dropped.

## III. Traffic Model

The proposed CAC scheme can be modeled by Markov Chain. Fig. 3 shows the Markov Chain for the queuing analysis of the overlaid macrocell layer, where the states of the system represent the number of calls in the system. The macrocell system provides $S$ number of additional states to support handover calls by the proposed adaptive QoS policy. The state $N$ is the maximum number of calls that can be accommodated by the macrocell system without QoS adaptation policy. Hence, the system provides QoS adaptation policy only to accept handover calls in the macrocell system. These handover calls include macrocell-to-macrocell and femtocell-to-macrocell handover calls. In Fig 3, the symbols $\lambda_{o,f}$ and $\lambda_{o,m}$, respectively, represents total originating call arrival rates considering all $n$ number of femtocells within a macrocell coverage area and only macrocell coverage area. $\lambda_{h,mm}$, $\lambda_{h,ff}$, $\lambda_{h,fm}$, and $\lambda_{h,mf}$, respectively, indicates the total macrocell-to-macrocell, femtocell-to-femtocell, femtocell-to-macrocell, and macrocell-to-femtocell handover call arrival rates within the macrocell coverage area.

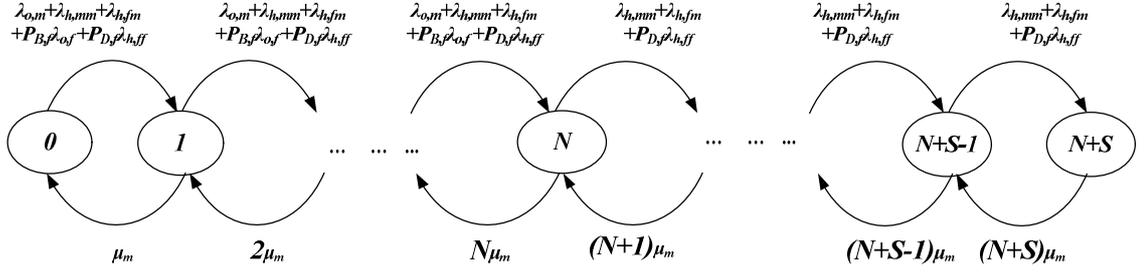

**Fig. 3.** The Markov Chain of a macrocell layer.

$P_{B,m}$ ($P_{B,f}$) is the new originating call blocking probability in the macrocell (femtocell) system. $P_{D,m}$ ($P_{D,f}$) is the handover call dropping probability in the macrocell (femtocell) system. The maximum number of calls that can be accommodated in a femtocell system is $K$. As the call arrival rate in a femtocell is very low and the data rate of a femtocellular network is high, there is no need of handover priority scheme for the femtocellular networks. The calls that are arrived in a femtocellular network are new originating calls, macrocell-to-femtocell handover calls, and the femtocell-to-femtocell handover calls. We define $\mu_m$ ($\mu_f$) as the channel release rate of macrocell (femtocell).

The average channel release rates for the femtocell layer and the macrocell layer are calculated as follows,

$$\mu_m = \eta_m(\sqrt{n}+1) + \mu \quad (1)$$

$$\mu_f = \eta_f + \mu \quad (2)$$

where $1/\mu$, $1/\eta_m$, and $1/\eta_f$ are, respectively, the average call duration (exponentially distributed), average cell dwell time for the macrocell (exponentially distributed), and the average cell dwell time for femtocell (exponentially distributed).

For the femtocell layer, the average call blocking probability, $P_{B,f}$ and the average call dropping probability, $P_{D,f}$ can be calculated as,

$$P_{D,f} = P_{B,f} = P_f(K) = \frac{\left(\frac{\lambda_{T,f}}{n}\right)^K \frac{1}{K!\mu_f^K}}{\sum_{i=0}^{K}\left(\frac{\lambda_{T,f}}{n}\right)^i \frac{1}{i!\mu_f^i}} \quad (3)$$

where $\lambda_{T,f} = \lambda_{f,o} + \lambda_{h,mf} + \lambda_{h,ff}$

The QoS adaptation/degradation policy is allowed for the handover calls of macrocell layer in our proposed scheme. For the macrocell layer, the average call blocking probability, $P_{B,m}$ and the average call dropping probability, $P_{D,m}$ can be calculated as,

$$P_{B,m} = \sum_{i=N}^{N+S} P(i) = \sum_{i=N}^{N+S} \frac{(\lambda_{m,0}+\lambda_{h,m})^N (\lambda_{h,m})^{i-N}}{i!\mu_m^i} P(0) \quad (4)$$

$$P_{D,m} = P(N+S) = \frac{(\lambda_{m,0}+\lambda_{h,m})^N \lambda_{h,m}^S}{(N+S)!\mu_m^{N+S}} P(0) \quad (5)$$

where $\lambda_{h,m} = \lambda_{h,mm} + \lambda_{h,fm} + P_{D,f}\lambda_{h,ff}$ and

$$P(0) = \left[\sum_{i=0}^{N} \frac{(\lambda_{m,0}+\lambda_{m,h})^i}{i!\mu_m^i} + \sum_{i=N+1}^{N+S} \frac{(\lambda_{m,0}+\lambda_{m,h})^N (\lambda_{m,h})^{i-N}}{i!\mu_m^i}\right]^{-1}$$

The handover call arrival rates are calculated as follows, The macrocell-to-macrocell handover call arrival rate is:

$$\lambda_{h,mm} = P_{h,mm} \frac{(1-P_{B,m})(\lambda_{m,o}+\lambda_{f,o}P_{B,f}) + (1-P_{D,m})(\lambda_{h,fm}+\lambda_{h,ff}P_{D,f})}{1-P_{h,mm}(1-P_{D,m})} \quad (6)$$

the macrocell-to-femtocell handover call arrival rate is:

$$\lambda_{h,mf} = P_{h,mf} \frac{(1-P_{B,m})(\lambda_{m,o}+\lambda_{f,o}P_{B,f}) + (1-P_{D,m})(\lambda_{h,fm}+\lambda_{h,ff}P_{D,f})}{1-P_{h,mm}(1-P_{D,m})} \quad (7)$$

the femtocell-to-femtocell handover call arrival rate is:

$$\lambda_{h,ff} = P_{h,ff} \frac{\lambda_{f,o}(1-P_{B,f}) + \lambda_{h,mf}(1-P_{D,f})}{1-P_{h,ff}(1-P_{D,f})} \quad (8)$$

and the femtocell-to-macrocell handover call arrival rate is:

$$\lambda_{h,fm} = P_{h,fm} \frac{\lambda_{f,o}(1-P_{B,f}) + \lambda_{h,mf}(1-P_{D,f})}{1-P_{h,ff}(1-P_{D,f})} \quad (9)$$

where $P_{h,mm}$, $P_{h,mf}$, $P_{h,ff}$, and $P_{h,fm}$ are, respectively, the macrocell-to-macrocell handover probability, macrocell-to-femtocell handover probability, femtocell-to-femtocell handover probability, and femtocell-to-macrocell handover probability.

The macrocell-to-macrocell handover probability, macrocell-to-femtocell handover probability, femtocell-to-femtocell handover probability, and femtocell-to-macrocell handover probability are calculated as follows,

The macrocell-to-macrocell handover probability is:

$$P_{h,mm} = \frac{\eta_m}{\eta_m + \mu} \quad (10)$$

the femtocell-to-macrocell handover probability is:

$$P_{h,fm} = \left[1 - n\left(\frac{r_f}{r_m}\right)^2\right]\frac{\eta_f}{\eta_f + \mu} \quad (11)$$

the femtocell-to-femtocell handover probability is:

$$P_{h,ff} = (n-1)\left(\frac{r_f}{r_m}\right)^2 \frac{\eta_f}{\eta_f + \mu} \quad (12)$$

and the macrocell-to-femtocell handover probability is:

$$P_{h,mf} = n\left(\frac{r_f}{r_m}\right)^2 \frac{\eta_m\sqrt{n}}{\eta_m\sqrt{n} + \mu} \quad (13)$$

where $r_f$ and $r_m$ are, respectively, the radius of a femtocell and radius of a macrocell coverage areas.

## IV. Performance Analysis

In this section we performed the effect of integrated macrocell/femtocell networks as well as the performance analysis of our proposed schemes. All the call arriving processes are assumed to be Poisson. The positions of the deployed femtocells within the macrocell coverage area are random. Table 1 shows the basic parameters that are used for the performance analysis. The propagation model for macrocell in [8] and propagation model for macrocell in [9] are used for the analysis.

**Table 1.** Summary of the parameter values used in analysis

| Parameter | Value |
|---|---|
| Radius of femtocell coverage area | 10 (m) |
| Carrier frequency for femtocells | 1.8 (GHz) |
| Transmit signal power by macrocellular BS | 1.5 KW |
| Maximum transmit power by a FAP | 10 (mW) |
| Height of macrocellular BS | 100 (m) |
| Height of FAP | 2 (m) |
| Height of MS | 2 (m) |
| Threshold value of received signal (RSSI) from a FAP to connect a call | -90 (dBm) |
| Bandwidth capacity of macrocell (C) | 6 (Mbps) |
| Required/allocated bandwidth for each of the QoS non-adaptive calls | 64 (Kbps) |
| Maximum required/allocated bandwidth for each of the QoS adaptive calls | 56 (Kbps) |
| Minimum required/allocated bandwidth for each of the QoS adaptive calls | 28 (Kbps) |
| Ratio of traffic arrivals (QoS non-adaptive calls: QoS adaptive calls) | 1:1 |
| Number of deployed femtocells in a macrocell coverage area | 1000 |
| Average call duration time (1/µ) considering all calls (exponentially distributed) | 120 (sec) |
| Average cell dwell time (1/η_f) for femtocell (exponentially distributed) | 360 (sec) |
| Average cell dwell time (1/η_m) for macrocell (exponentially distributed) | 240 (sec) |
| Density of call arrival rate (at femtocell coverage area: at macrocell only coverage area) | 20:1 |
| Standard deviation for the lognormal shadowing loss | 8 (dB) |
| Penetration loss | 20 (dB) |

Fig. 4 shows the effect of different handover probabilities with the increase of the number of deployed femtocells within a macrocellular network coverage. With the increase of the number of deployed femtocells, the femtocell-to-femtocell handover and macrocell-to-femtocell handover probabilities are significantly increased. Also, the femtocell-to-macrocell handover probability is very high. Therefore admission control is the important issue for the dense femtocellular network deployment. Whenever the macrocell and the femtocells are integrated, a huge number of macrocell calls are diverted to femtocells through the macrocell-to-femtocell handover calls. As a consequence, the macrocell system is able to accommodate more number of calls. Therefore, the integrated macrocell/femtocell networks significantly reduces the overall forced call termination probability in the macrocellular networks. Fig. 5 shows the performance improvement of macrocellular networks in terms of overall forced call termination probability. Hence, the proposed CAC is able to accept huge amount of handover calls as well as new calls in the system. The improvement of the overall forced call termination performance increases the revenue for the operators.

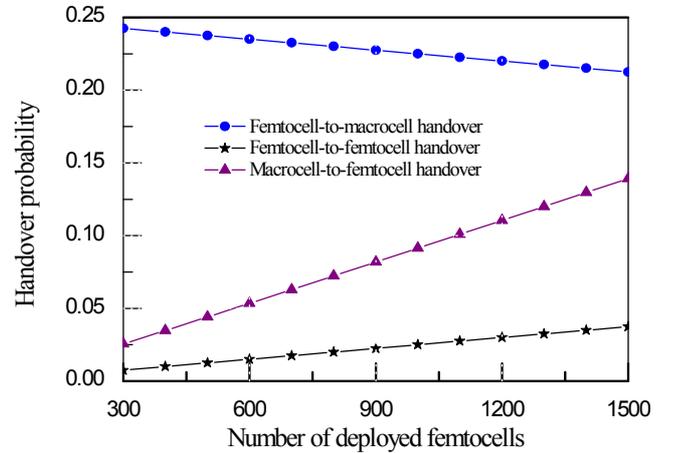

**Fig. 4.** Handover probability comparison.

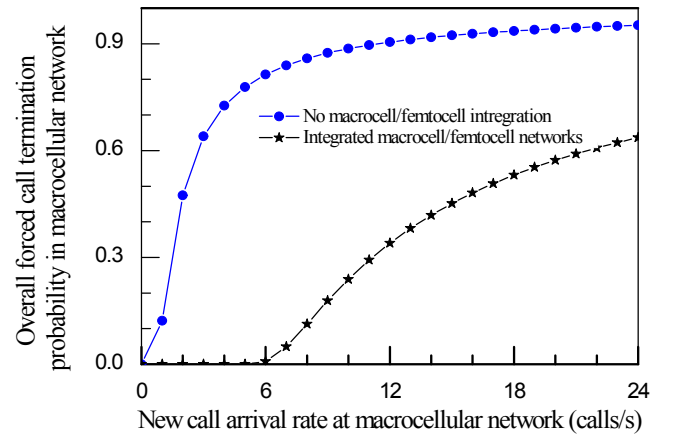

**Fig. 5.** Overall forced call termination probability in the macrocell system for two different cases.

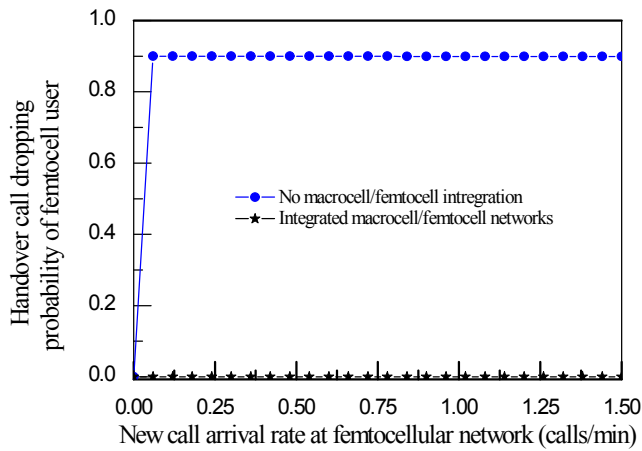

**Fig. 6.** Comparison of handover call dropping probability of femtocell users.

Due to movement of the users, the femtocell users move inside and outside the serving femtocell coverage area. If there is no integrated macrocell/femtocell network system, femtocell-to-macrocell handover calls will be surely dropped. As a result the handover call dropping probability for the femtocell users will be very high. Fig. 6 shows that the integrated macrocell/femtocell networks result in negligible handover call dropping probability for the femtocell users.

As the number of deployed femtocells is increased, the femtocell-to-macrocell handover call arrival rate is increased. These increased number of handover calls are supported by the QoS adaptability policy in macrocellular networks. The results in Fig. 4 – Fig. 6 show the performance improvement of the proposed scheme. The proposed QoS adaptive/degradation policy is able to handle huge number of handover calls.

## V. Conclusions

Femtocell technology is one of the most promising supporting technologies for the 4G/5G wireless communication. It holds the potential of increasing the cellular network capacity by offloading some of the traffic of the macrocellular installations. The integrated femtocell/macrocell network is the attractive solution for the future convergence networks. It can provide higher QoS for indoor users at low price, while simultaneously reducing the burden on the whole network system. The suggested CAC policy is able to manage huge handover calls inside a macrocell coverage. The traffic model for a macrocell/femtocell integrated network is very effective for the performance analysis of a macrocell/femtocell integrated network. The results demonstrated in this paper obviously indicate the advantages of our proposed schemes.

## Acknowledgments

This work was supported by the IT R&D program of MKE/KEIT [10035362, Development of Home Network Technology based on LED-ID].